\documentclass[prl,twocolumn,tbtags,showpacs,superscriptaddress,floatfix]{revtex4}

\usepackage{graphicx,amsmath}
\usepackage{bm}

\newcommand{\NLij}{i\kern -0.08em j}
\newcommand{\half}{{\textstyle\frac{1}{2}}}

\newcommand{\ua}{\uparrow}
\newcommand{\da}{\downarrow}
\newcommand{\ket}[1]{\lvert{#1}\rangle}

\begin{document}

\title{Controllable coupling of superconducting flux qubits}

\author{S.H.W. van der Ploeg}
\email{simon.vanderploeg@ipht-jena.de}
\thanks{SvdP and AI contributed equally to this work.}
\affiliation{%
Institute for Physical High Technology, P.O. Box 100239, D-07702 Jena, Germany}

\author{A.~Izmalkov}
\affiliation{%
Institute for Physical High Technology, P.O. Box 100239, D-07702 Jena, Germany}

\author{Alec \surname{Maassen van den Brink}}
\email{alec@riken.jp}
\affiliation{%
D-Wave Systems Inc., 100-4401 Still Creek Drive, Burnaby, B.C., V5C
6G9 Canada} \affiliation{Frontier Research System, RIKEN, Wako-shi,
Saitama, 351-0198, Japan}

\author{U.~H\"ubner}
\affiliation{%
Institute for Physical High Technology, P.O. Box 100239, D-07702 Jena, Germany}

\author{M.~Grajcar}
\affiliation{%
Department of Solid State Physics, Comenius University, SK-84248 Bratislava, Slovakia}

\author{E.~Il'ichev}
\affiliation{%
Institute for Physical High Technology, P.O. Box 100239, D-07702 Jena, Germany}

\author{H.-G.~Meyer}
\affiliation{%
Institute for Physical High Technology, P.O. Box 100239, D-07702 Jena, Germany}

\author{A.M. Zagoskin}
\email{zagoskin@riken.jp} \affiliation{Frontier Research System,
RIKEN, Wako-shi, Saitama, 351-0198, Japan}
\affiliation{%
Physics and Astronomy Dept., The University of British Columbia,
6224 Agricultural Rd., Vancouver, B.C., V6T 1Z1 Canada}

\date{\today}

\begin{abstract}
We have realized controllable coupling between two three-junction flux qubits by inserting an additional coupler loop between them, containing three Josephson junctions. Two of these are shared with the qubit loops, providing strong qubit--coupler interaction. The third junction gives the coupler a nontrivial current--flux relation; its derivative (i.e., the susceptibility) determines the coupling strength $J$, which thus is tunable \emph{in situ} via the coupler's flux bias. In the qubit regime, $J$ was varied from $\sim$45 (antiferromagnetic) to $\sim${}$-55$~mK (ferromagnetic); in particular, $J$~vanishes for an intermediate coupler bias. Measurements on a second sample illuminate the relation between two-qubit tunable coupling and three-qubit behavior.
\end{abstract}


\maketitle

The development of Josephson qubit devices has led to the realization of quantum gates \cite{Pashkin,Yamamoto,Chiorescu} as well as two- \cite{Pashkin,Berkley,Izmalkov} and four-qubit~\cite{4qb} coupling. For the implementation of real quantum algorithms, several hurdles must still be cleared, such as increasing the number of qubits and their coherence times. Equally important, however, is coupling \emph{tunability}. If the coupling strength can be continuously tuned between two values with opposite signs, it can be naturally switched off---a great advantage when applying two-qubit gates~\cite{untun}. Moreover, in adiabatic quantum computing, continuously tuning the Hamiltonian is crucial, and both ferromagnetic (FM) and antiferromagnetic (AF) couplings are necessary~\cite{AQC_SC}. In a promising group of proposals, coupling capacitances and inductances are replaced with \emph{effective} (``quantum'') capacitances~\cite{AB} and inductances \cite{Clarke_coupler,Alec_Berkley}, respectively, which are (sign-)\linebreak[1]tunable via their bias dependence.

\begin{figure}[tb]
\includegraphics[width=7.5 cm,angle=0]{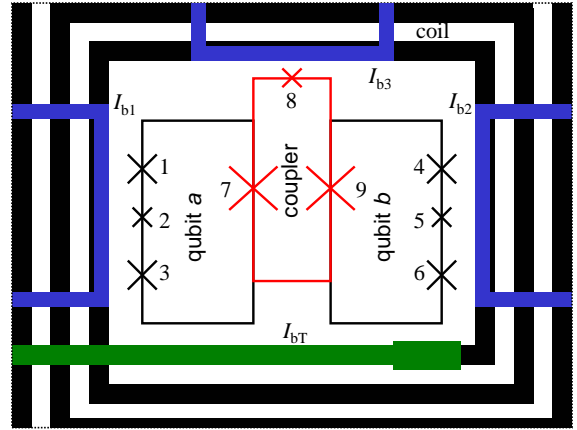}
\caption{Circuit design of sample~1. Junctions 123 (456) form
qubit~\textit{a} (\textit{b}). The large qubit junctions have areas
$S_{1,3,4,6}=150\times700$~nm$^2$, while $\alpha_{a,b}\equiv
S_{2,5}/S_1=0.65$. Junctions 789 form the coupler (red), with areas
$S_{7,9}=150\times2000$~nm$^2$ and $S_8=150\times400$~nm$^2$. The
ratio between the coupler and qubit loop areas is $1:2$. The tank
circuit has inductance $L=90$~nH and capacitance $C=470$~pF,
yielding a resonance at $\omega_\mathrm{T}/2\pi=20.76$~MHz with
quality $Q=300$. The mutual inductances, deduced from the flux
periodicity, are $M_{a,\mathrm{T}}=M_{b,\mathrm{T}}=98$,
$M_\mathrm{c,T}=51$, $M_{a,\mathrm{b1}}=0.85$,
$M_{b,\mathrm{b1}}=0.17$, $M_\mathrm{c,b1}=0.19$,
$M_{a,\mathrm{b2}}=0.23$, $M_{b,\mathrm{b2}}=1.2$,
$M_\mathrm{c,b2}=0.25$, $M_{a,\mathrm{b3}}=0.42$,
$M_{b,\mathrm{b3}}=0.39$, and $M_\mathrm{c,b3}=0.56$ (all in pH).}
\label{fig1}
\end{figure}

We report the realization of \emph{sign}-tunable coupling~\cite{Harris} between three-Josephson-junction (3JJ) flux qubits \cite{Mooij}. These have a low charge-noise sensitivity, common to all flux qubits. Their small area also protects them reasonably well against magnetic noise, but limits the strength of their AF coupling via magnetic~\cite{Izmalkov} and/or kinetic~\cite{majer} inductance. This can be overcome by using a \emph{Josephson} mutual inductance~\cite{Delft_ladder}, which can also be ``twisted'' for FM coupling or (in theory) current-biased for limited tunability~\cite{Coupling2005}. Our design~\cite{Alec_note} combines the above ideas, mediating a tunable galvanic coupling through a ``quantum Josephson mutual inductance''. The coupler is connected to qubits $a$ and~$b$ via shared junctions 7 and~9, see Fig.~\ref{fig1}. By changing the coupler's flux bias $\Phi_\mathrm{c}^\mathrm{x}=f_\mathrm{c}\Phi_0$ ($\Phi_0$ is the flux quantum), the phase difference across junction~8 and therefore the interaction strength~$J$ can be tuned. The fluxes through the coupler and qubits are controlled by bias-line currents $I_\mathrm{b1,2,3}$ and the dc component~$I_\mathrm{bT}$ of the coil current~\cite{compensation}.

The system can be described by the effective pseudospin Hamiltonian \cite{Mooij,Izmalkov,Chiorescu}
\begin{equation}
 H= -\sum_{i=a,b} [\epsilon_i \sigma_z^{(i)} + \Delta_i
 \sigma_x^{(i)}] + J(\phi_\mathrm{c})\sigma_z^{(a)}\sigma_z^{(b)}\;, \label{H}
\end{equation}
where $\epsilon_i$ is the bias on qubit~$i$, $\Delta_i$ is the corresponding tunnelling matrix element, $\sigma_z^{(i)}$, $\sigma_x^{(i)}$ are Pauli matrices in the flux basis, and $\phi_\mathrm{c}\equiv2\pi f_\mathrm{c}$.

\begin{figure}[t]
\includegraphics[width=8cm]{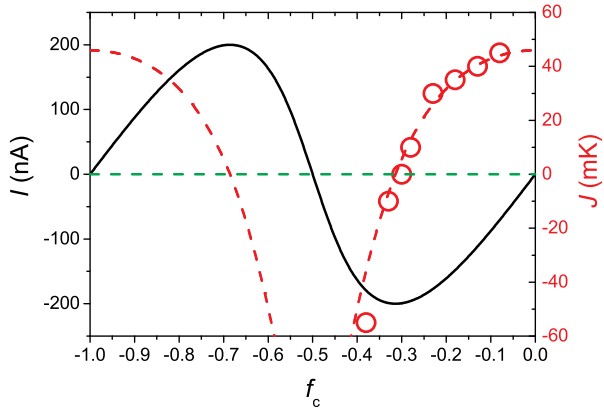}
\caption{Black: the current--flux relation $I(f_\mathrm{c})$ of a
coupler with $\alpha_\mathrm{c}=S_8/S_9=0.2$ and
$I_\mathrm{c}=1~\mu$A. Red-dashed line: the coupling energy
$J(f_\mathrm{c})$ obtained from Eq.~(\ref{J-res}) using this
$I(f_\mathrm{c})$ and the loop currents $I_{\mathrm{p}a,b}$ found
independently from the qubit response. Circles: experimental
$J(f_\mathrm{c})$ obtained from Fig.~\ref{fig3}.} \label{fig2}
\end{figure}

\begin{figure*}[t]
\includegraphics[scale=0.24]{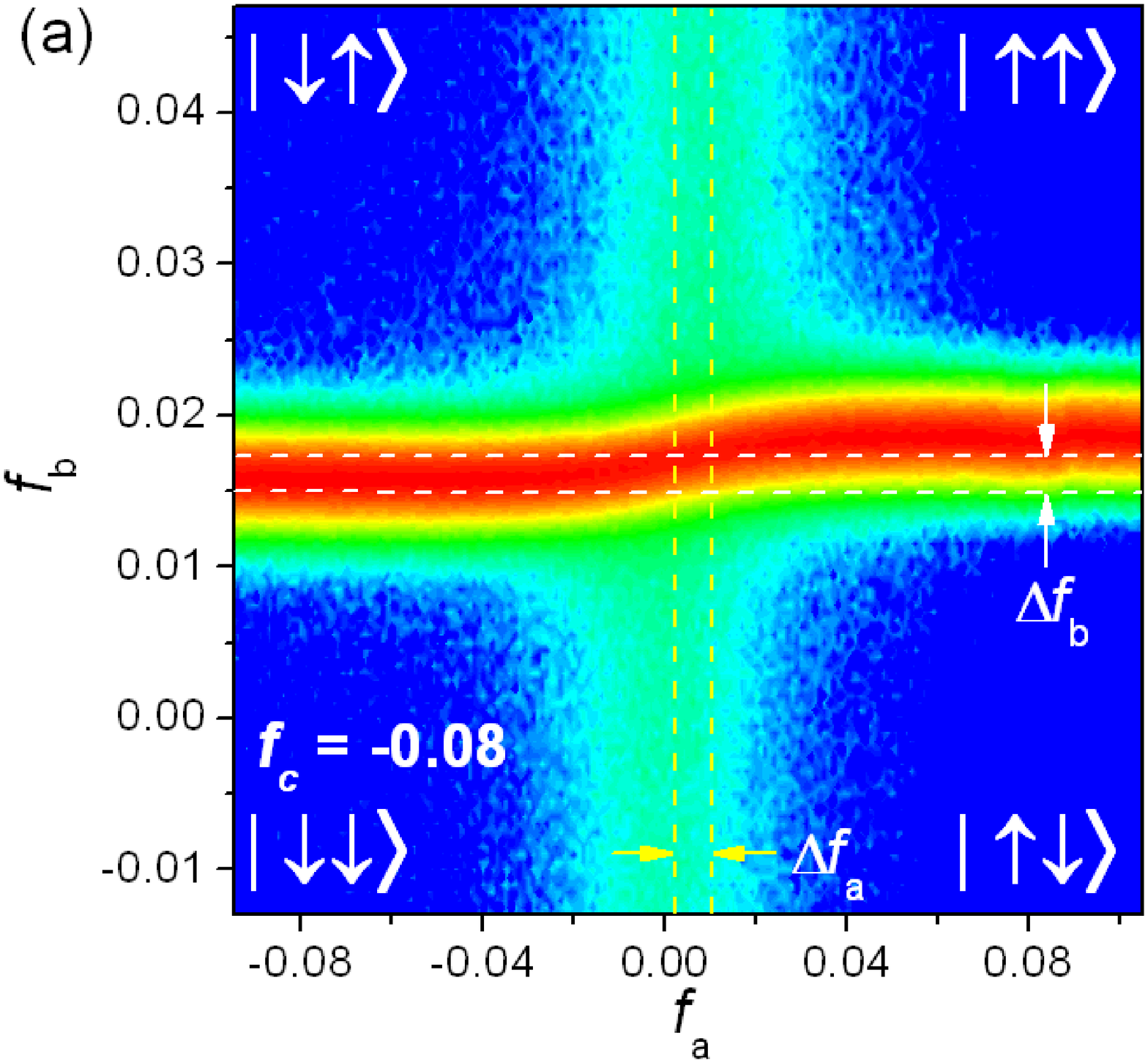}\hfill\includegraphics[scale=0.24]{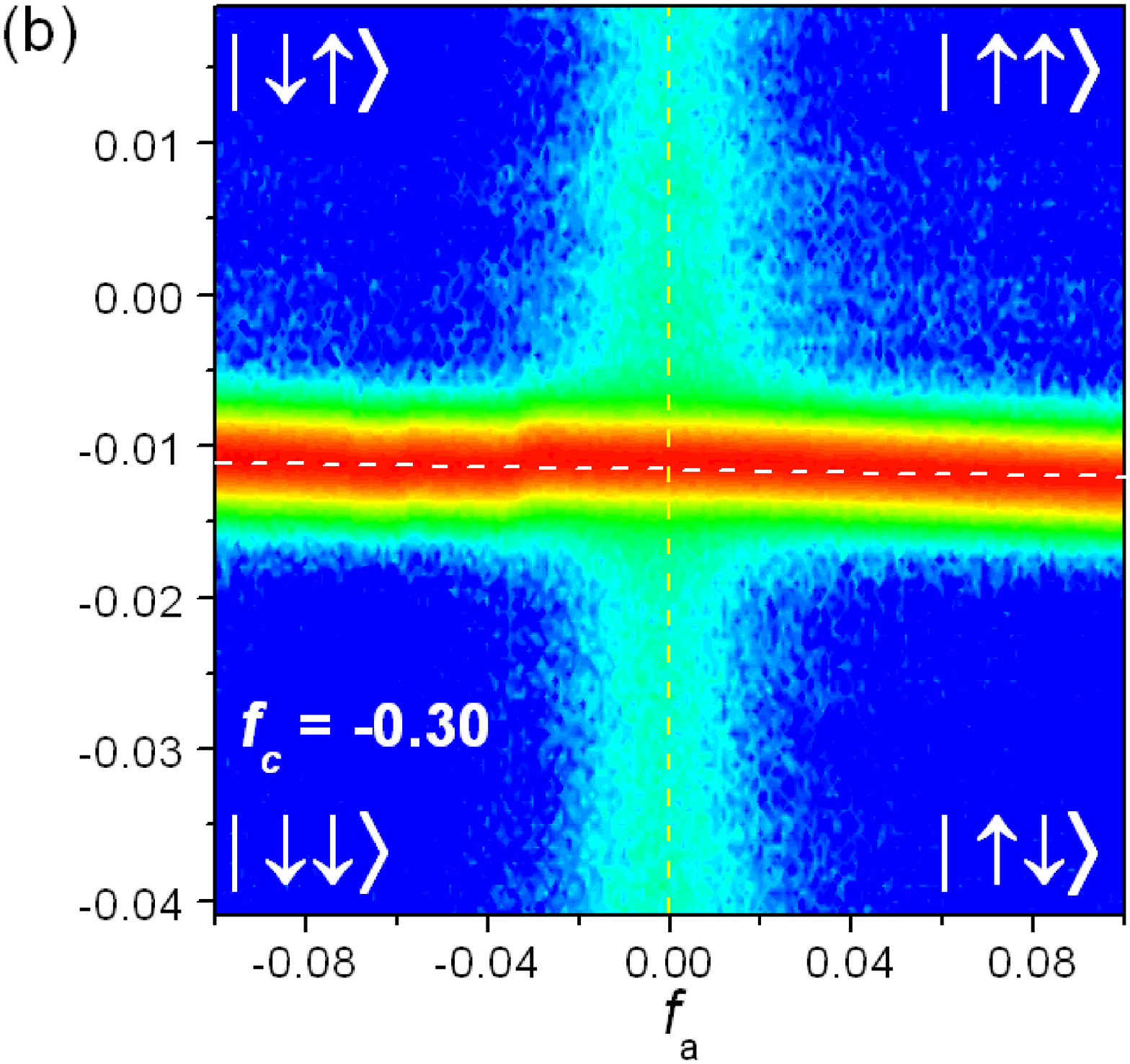}\hfill\includegraphics[scale=0.24]{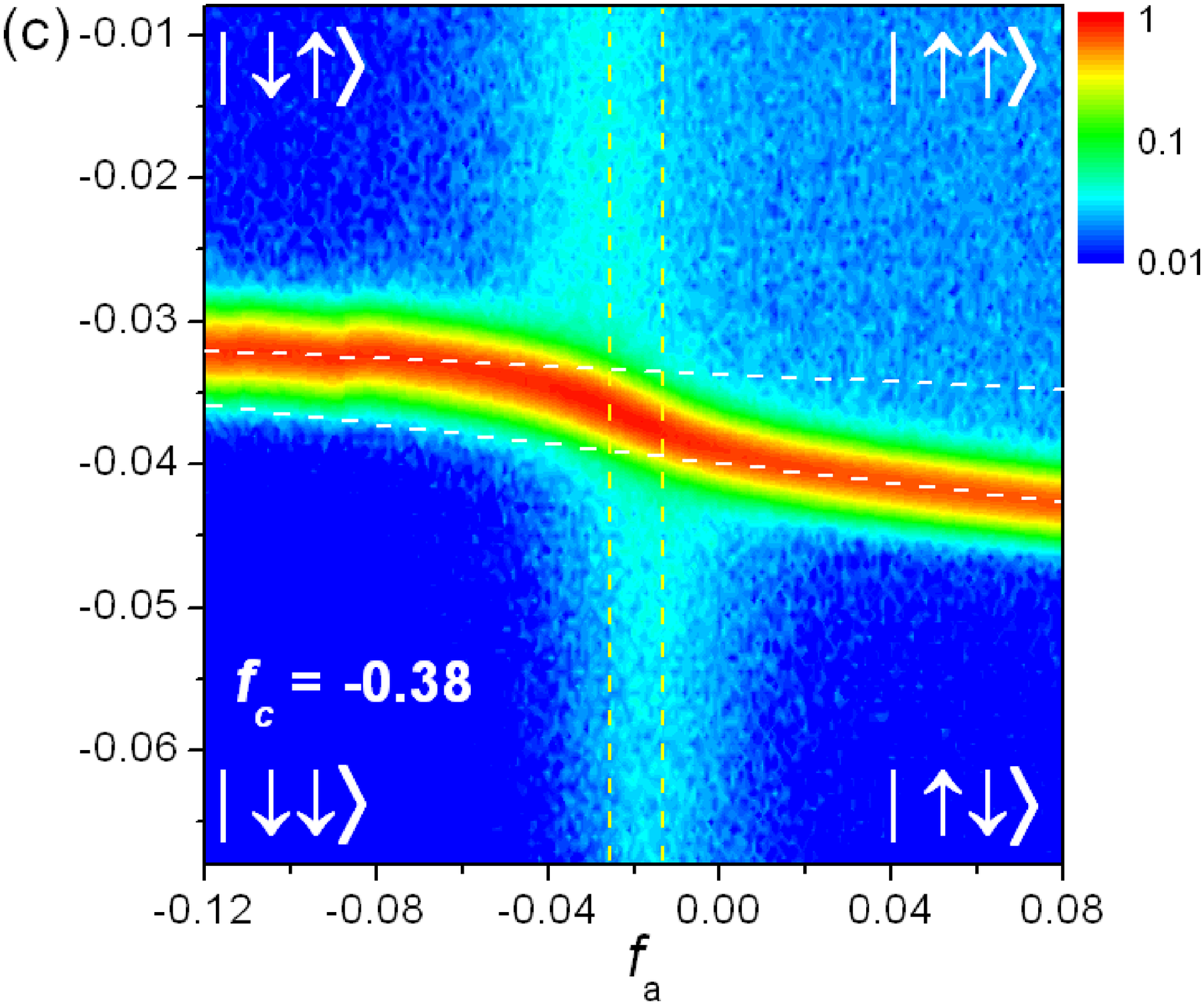}
\caption{$-\tan\theta(f_a,f_b)$ for sample~1 at coupler bias
$f_\mathrm{c}=-0.08$, $-0.30$, and $-0.38$, with a manifest change
in coupling sign. A theory fit as in Fig.~\ref{fig4} yields
couplings $J=45$, $0$, and $-55$~mK. The excess response in the
$\ket{\ua\ua}$ quadrant for $f_\mathrm{c}=-0.38$ is due to the
coupler.} \label{fig3}
\end{figure*}

To calculate $J$ for a coupler with symmetric Josephson energies $E_{7,9}=E$, $E_8=\alpha_\mathrm{c}E$, consider the potential
\begin{equation}\begin{split}
  U_\mathrm{J}(\bm{\phi})&=
  -E[\cos\phi_7+\alpha_\mathrm{c}\cos(\phi_\mathrm{c}{-}\phi_7{-}\phi_9)+\cos\phi_9]\\
  &\quad\,+U_a(\phi_a{+}\phi_7)+U_b(\phi_b{+}\phi_9)\;,
  \label{UJ}
\end{split}\end{equation}
where we implemented flux quantization for small-inductance loops,
with $\phi_{a,b}$ being the qubit flux biases in phase units, and
where the qubit energies $U_{a,b}$ are already minimized over their
internal degrees of freedom~$\phi_\text{1--6}$. However, each qubit
has two minimum states, with opposite values of the persistent
currents $-(2e/\hbar)U_{a,b}'=\pm I_{\mathrm{p}a,b}$. We minimize
$U_\mathrm{J}(\bm{\phi})$ with respect to $\phi_{7,9}$, and expand
in~$E^{-1}$. This implements the classical limit of the general
condition that the coupler should stay in its ground state,
following the qubits adiabatically~\cite{AB}. To leading order, the
phases obey $\phi_7=\phi_9=\bar{\phi}$, with
$\sin\bar{\phi}=\alpha_\mathrm{c}\sin(\phi_\mathrm{c}{-}2\bar{\phi})$.
Proceeding to $\mathcal{O}(E^{-1})$ and retaining terms $\propto
I_{\mathrm{p}a}I_{\mathrm{p}b}$, one finds~\cite{J-note}
\begin{equation}
  J(\phi_\mathrm{c})=\frac{\hbar}{2e}
  \frac{I'(\phi_\mathrm{c})}{I_\mathrm{c}^2-I(\phi_\mathrm{c})^2}\,I_{\mathrm{p}a}I_{\mathrm{p}b}\;,
  \label{J-res}
\end{equation}
in terms of the coupler current $I(\phi_\mathrm{c})=I_\mathrm{c}\sin\bar{\phi}$ with $I_\mathrm{c}=(2e/\hbar)E$, so that
\begin{equation}
  I'(\phi_\mathrm{c})=I_\mathrm{c}
  \frac{\cos(\bar{\phi})\alpha_\mathrm{c}\cos(\phi_\mathrm{c}{-}2\bar{\phi})}
       {\cos(\bar{\phi})+2\alpha_\mathrm{c}\cos(\phi_\mathrm{c}{-}2\bar{\phi})}\;.
  \label{I-slope}
\end{equation}
The numerator in Eq.~(\ref{J-res}) also occurs for magnetic coupling~\cite{Alec_Berkley}; the denominator reflects, for finite coupler-loop currents, the nonlinearity of the Josephson elements 7 and~9. Figure~\ref{fig2} shows $I$ and $J$ for $\alpha_\mathrm{c}=0.2$. If $\alpha_\mathrm{c}\ll\nobreak1$, $I(\phi_\mathrm{c})\approx\alpha_\mathrm{c}I_\mathrm{c}\sin\phi_\mathrm{c}$, i.e.,
$J(\phi_\mathrm{c})\approx\linebreak[1] I_{\mathrm{p}a}I_{\mathrm{p}b}\*\hbar^2\alpha_\mathrm{c}\cos\phi_\mathrm{c}/4e^2E$~\cite{Alec_note}. Hence, $J(\phi_\mathrm{c})>0$ ($<\nobreak0$) near $\phi_\mathrm{c}=0$ ($\pi$)~\cite{extreme}, corresponding to AF (FM) coupling. However, Fig.~\ref{fig2} already is strongly non-sinusoidal, with a larger maximum for FM than for AF coupling (cf.~Ref.~\cite{Alec_Berkley}).

The qubit--coupler circuit was fabricated out of aluminum, and the pancake coil out of niobium \cite{Izmalkov,4qb}. Besides providing an overall dc field bias, the coil is part of an $LC$ tank circuit driven at resonance, well below the characteristic qubit frequencies. In the Impedance Measurement Technique~\cite{Grajcar}, one records the tank's current--voltage phase angle~$\theta$, which is very sensitive to its effective inductance: $\tan\theta=-(Q/L)\chi'$, where $Q$ and $L$ are the tank's quality factor and ``bare'' inductance, respectively, and $\chi'$ is the contribution to the tank inductance due to the qubits' reactive susceptibility. In the coherent regime, $\chi'$ has peaks at level anticrossings~\cite{greenberg}, where a small bias $\epsilon_i\sim\Delta_i$ will flip the flux state of qubit~$i$. Thus, these peaks demarcate the qubits' stability diagram, from which $J$ can be read off~\cite{Coupling2005}, while their widths are $\propto\Delta_i$.

\begin{figure}[hbp]
\includegraphics[scale=0.24]{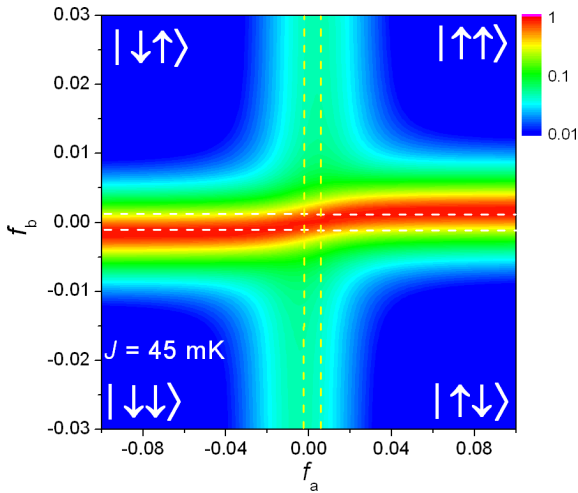}
\caption{Theoretical fit for Fig.~\ref{fig3}a. The extracted
parameters are $T_\mathrm{eff}=70$~mK, $\Delta_a=300$~mK,
$I_{\mathrm{p}a}=75$~nA, $\Delta_b=55$~mK, $I_{\mathrm{p}b}=180$~nA,
and $J(-0.08)\approx45$~mK.} \label{fig4}
\end{figure}

All measurements were performed in a dilution refrigerator with a base temperature of 10~mK. Results for sample~1 are presented in Fig.~\ref{fig3} around the qubits' co-degeneracy point [$\epsilon_a=\epsilon_b=0$ in~(\ref{H})], for fixed~\cite{compensation} coupler fluxes $f_\mathbf{c}= -0.08$, $-0.30$, and $-0.38$. In Fig.~\ref{fig3}a, the FM ordered $\ket{\ua\ua}$ and $\ket{\da\da}$ states are pushed away from this point, where the $\ket{\ua\da}$ and $\ket{\da\ua}$ states dominate---a clear signature of AF coupling. In Fig.~\ref{fig3}b, the $a$-~(vertical) and $b$\nobreakdash-traces (horizontal) are independent, demonstrating zero coupling. Finally, Fig.~\ref{fig3}c is opposite to Fig.~\ref{fig3}a, corresponding to FM coupling.

Quantitatively, the state of the system~(\ref{H}) at temperature $T_\mathrm{eff}$ is readily calculated; the effect on the tank flux follows from the mutual inductances as in Fig.~\ref{fig1}, and taking the $I_\mathrm{bT}$\nobreakdash-derivative yields~$\chi'$. Fitting this equilibrium response to the data, all parameters in Eq.~(\ref{H}) as well as $T_\mathrm{eff}$ can be extracted \cite{Izmalkov,4qb,Coupling2005,Grajcar}. From the shape of the single-qubit traces, one finds $\Delta_{a,b}$ and~$I_{\mathrm{p}a,b}$; the shifts $\Delta f_{a,b}$ in these traces when they cross the co-degeneracy point yield $J=\frac{1}{2}\Delta f_{a,b} \Phi_0I_{\mathrm{p}a,b}$. That is, here we foremost measure~$H$, not the qubit state~$\ket{\Psi}$. As an example, Fig.~\ref{fig4} shows a fit to Fig.~\ref{fig3}a, yielding $T_\mathrm{eff}\approx70$~mK, $\Delta_a=300$~mK, $I_{\mathrm{p}a}=75$~nA, $\Delta_b=55$~mK, $I_{\mathrm{p}b}=180$~nA, and $J \approx 45$~mK. The agreement between theory and experiment confirms that the system is in the qubit regime.

The $J(f_\mathrm{c})$ thus measured (Fig.~\ref{fig2}) agrees with Eq.~(\ref{J-res}) for the design value $\alpha_\mathrm{c}=0.2$, and $I_\mathrm{c}=1~\mu$A for the coupler. Note that $J(f_\mathrm{c}{=}{-}0.30)\linebreak[1]=0$ already implies $\alpha_\mathrm{c}\approx0.2$ by Eqs.\ (\ref{J-res}) and~(\ref{I-slope}); $I_\mathrm{c}=1~\mu$A is consistent with $I_\mathrm{c} \approx\linebreak[1](S_9/S_6)\*I_{\mathrm{p}b}/\sqrt{1-1/4\alpha^2} \approx 0.8~\mu$A, expected from~$I_{\mathrm{p}b}$ ($S_6$, $S_9$ are the respective junction areas).


\begin{figure}[hbp]
\includegraphics[scale=0.24]{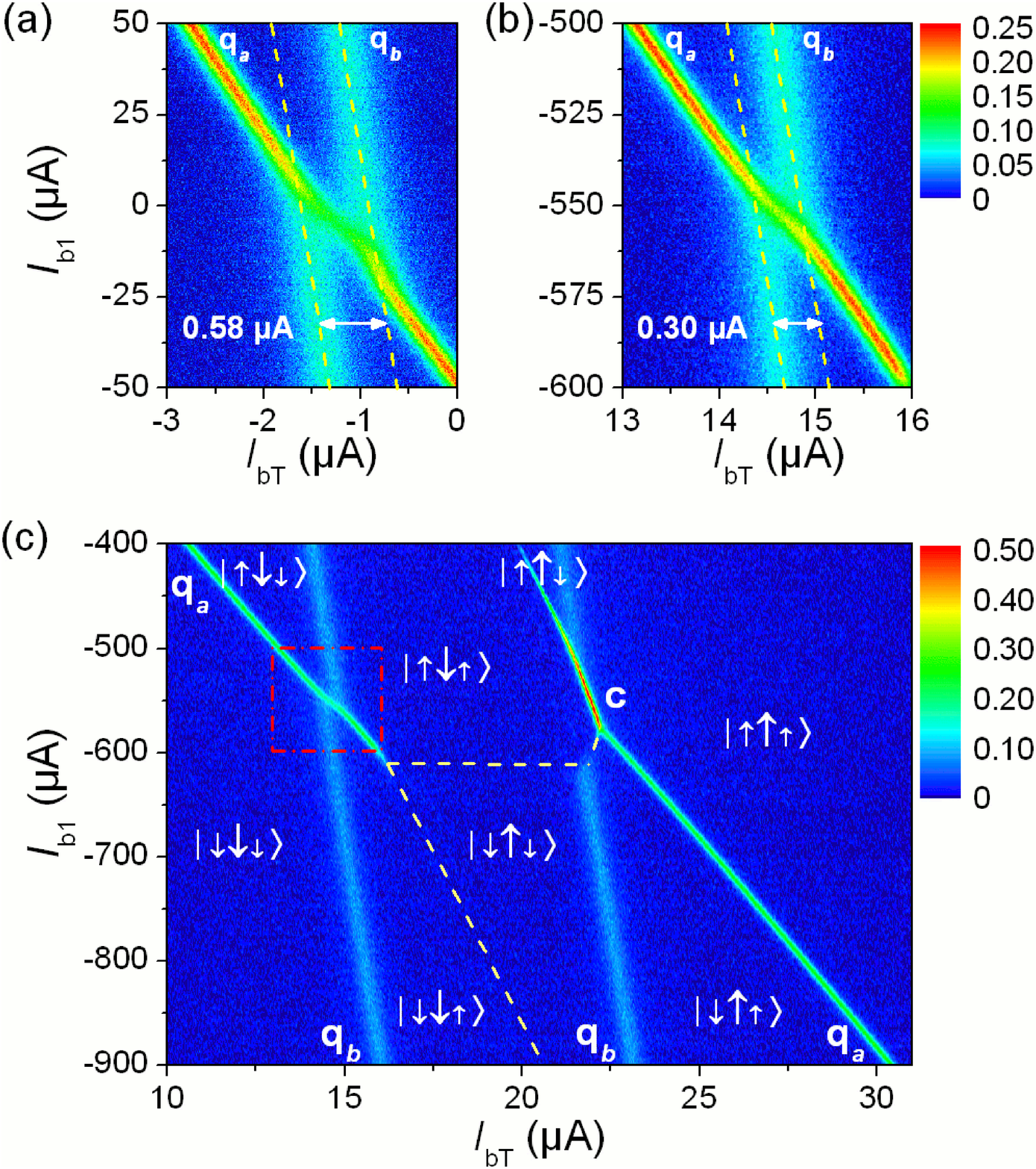}
\caption{$-\tan\theta(I_\mathrm{bT},I_\mathrm{b1})$ for sample~2.
(a)~$I_\mathrm{b2}=0$. At co-degeneracy, the coupler flux is
$f_\mathrm{c}=0.237$, and the induced coupling $J\approx85$~mK.
(b)~and (c) $I_\mathrm{b2}=450$~$\mu$A. (b)~Close-up of the
co-degeneracy point [dashed box in~(c)], where $f_\mathrm{c}=0.347$
and $J\approx44$~mK. (c)~Overview and stability diagram. At bias
points outside the boxed area, the coupler itself can become
bistable, and contribute its own anticrossing to the impedance
response. The arrow sizes denote the relative magnitudes of the loop
currents in qubit~$a$, coupler, and qubit~$b$ respectively.}
\label{fig5}
\end{figure}

A sample~2 was fabricated with $\alpha_\mathrm{c}=0.5$, i.e., a larger junction~8 and hence potentially a stronger coupling. Figure~\ref{fig5} presents results for $\tan\theta(I_\mathrm{bT},I_\mathrm{b1})$. In Fig.~\ref{fig5}a, $I_\mathrm{b2}=0$. The qubits are at co-degeneracy for $f_\mathrm{c}=0.237$. A theoretical fit as for sample~1 gives $I_{\mathrm{p}a}=130$~nA, $I_{\mathrm{p}b}=70$~nA, $\Delta_a\approx\Delta_b\approx70$~mK, and $J\approx85$~mK. In Fig.~\ref{fig5}b, a bias current $I_\mathrm{b2}=450~\mu$A is applied. As a result, the co-degeneracy point is shifted to $f_\mathrm{c}=0.347$. Compared with Fig.~\ref{fig5}a, the coupling strength is reduced to $J=85\times(0.30/0.58)\approx44$~mK.

For $\alpha_\mathrm{c}=0.5$, the negative-slope portion of $I(f_\mathrm{c})$ (cf.\ Fig.~\ref{fig2}) is very narrow, so that FM coupling should only occur for $|f_\mathrm{c}{-}\half|\ll1$. Hence, in sample~2 only the (AF) coupling strength could be varied, not its sign. In fact, the coupler is on the very boundary of the hysteretic regime~\cite{Mooij}. Hence, for $f_\mathrm{c}\approx\half$, the energy gap above the ground state will be very small and the adiabaticity condition mentioned above Eq.~(\ref{J-res}) breaks down, regardless of the exact value of $\alpha_\mathrm{c}-\half$. Indeed, in this case we rather observe \emph{three}-qubit behavior, with the coupler's own anticrossing characterized by $\Delta_\mathrm{c}=90$~mK and $I_\mathrm{pc}=300$~nA (Fig.~\ref{fig5}c). This is fully consistent with our interpretation above, since the operating regimes are different.

In conclusion, we have for the first time demonstrated sign-tunable
Josephson coupling between two three-junction flux qubits, in the
quantum regime. At $T_\mathrm{eff}\approx70$~mK, the coupling
strength $J$ was changed from +45 (antiferromagnetic) to $-55$~mK
(ferromagnetic). At an intermediate coupler bias, $J$~vanishes,
thereby realizing the elusive superconducting switch. These results
represent considerable progress towards solid-state quantum
computing in general. The present low-frequency mode of operation is
particularly attractive for adiabatic quantum computing: control of
$J>0$ is necessary to operate the computer, and sufficiently strong
ferromagnetic coupling ($J<0$) allows one to create dummy qubits, as
used in the scalable architecture of Ref.~\cite{AQC_SC}. While our
measurements are essentially equilibrium, the design of
Fig.~\ref{fig1} is also relevant in the ac domain, where the
coupling can be controlled by a resonant rf signal~\cite{ac-tune}.

AMvdB thanks M.H.S. Amin, J.M. Martinis, and A.Yu.\ Smirnov for
discussions, and the ITP (Chinese Univ.\ of Hong Kong) for its
hospitality. SvdP, AI, and EI were supported by the EU through the
RSFQubit and EuroSQIP projects, MG by Grants VEGA 1/2011/05 and
APVT-51-016604 and the Alexander von Humboldt Foundation, and AZ by
the NSERC Discovery Grants Program.

\emph{Note added.}---Recently, we learned of the work of Hime
\emph{et al.} \cite{Hime} implementing the controllable-coupling
proposal described in Ref.\cite{Clarke_coupler}.


\begin{thebibliography}{99}
\bibitem{compensation} For sample~1, $I_\mathrm{bT}$ is used to set the overall magnetic field, while $I_\mathrm{b1,2,3}$ in combination facilitate independent tuning of the biases in qubits and coupler using a flux-compensation procedure. Sample~2 does not have a wire $I_\mathrm{b3}$ and is measured without flux compensation.
\bibitem{J-note} Since corrections are $\mathcal{O}[(I_{\mathrm{p}a,b}/I_\mathrm{c})^2]$, with $I_{\mathrm{p}a,b}\sim\nobreak\frac{1}{4}I_\mathrm{c}$, Eq.~(\ref{J-res}) should be a reasonable approximation.
\end{thebibliography}
\end{document}